\documentclass{article}
\usepackage[utf8]{inputenc}
\usepackage{graphicx} 
\usepackage{cprotect}
\usepackage{setspace}
\usepackage[top=1in, bottom=1in, right=1in, left=1in]{geometry}
\usepackage[colorlinks,allcolors=blue]{hyperref}
\usepackage[round]{natbib}
\usepackage{amsmath,amssymb,verbatim,amsthm,xfrac,bbm,mathabx}

\newsavebox\verbbox

\begin{lrbox}{\verbbox}
\footnotesize \verb+https://github.com/alexandercoulter/fastfrechet+
\end{lrbox}

\begin{document}

\onehalfspacing

\begingroup
\centering
\Large \verb+fastfrechet+: \textbf{An R package for fast implementation of Fréchet regression with distributional responses}\\
\normalsize
\begin{table}[h!]
    \centering
    \begin{tabular}{l|l}
    \textbf{Alexander Coulter} & \textit{Department of Statistics, Texas A\&M University}\\
    \textbf{Rebecca Lee} & \textit{Department of Statistics, Texas A\&M University}\\
    \textbf{Irina Gaynanova} & \textit{Department of Biostatistics, University of Michigan}
\end{tabular}
\end{table}
\today\par
\endgroup

\section{Introduction}

Distribution-as-response regression problems are gaining wider attention, especially within biomedical settings where observation-rich patient specific data sets are available, such as feature densities in CT scans~\citep{petersen_wasserstein_2021} actigraphy~\citep{ghosal_distributional_2023}, and continuous glucose monitoring~\citep{coulter_fast_2024, matabuena_glucodensities_2021}. To accommodate the complex structure of such problems,~\citet{petersen_frechet_2019} proposed a regression framework called \textit{Fréchet regression} which allows non-Euclidean responses, including distributional responses. This regression framework was further extended for variable selection by~\citet{tucker_variable_2023}, and~\citet{coulter_fast_2024} developed a fast variable selection algorithm for the specific setting of univariate distributional responses equipped with the 2-Wasserstein metric (\textit{2-Wasserstein space}). We present \verb+fastfrechet+\footnote{Available at \usebox{\verbbox}, with planned release on the Comprehensive R Archive Network (CRAN). Currently supported only on MacOS and Windows.}, an \textsc{R} package providing fast implementation of these Fréchet regression and variable selection methods in 2-Wasserstein space, with resampling tools for automatic variable selection. \verb+fastfrechet+ makes distribution-based Fréchet regression with resampling-supplemented variable selection readily available and highly scalable to large data sets, such as the UK Biobank~\citep{doherty_large_2017}.

Fréchet regression with variable selection is currently not implemented by any software package, available only through the Supplementary Material of \citet{tucker_variable_2023} (hereafter ``\verb+Tucker+ materials"). As discussed in \citet{coulter_fast_2024}, the implemented algorithm is prohibitively slow. Implementation of Fréchet regression in 2-Wasserstein space without variable selection is supported by the \verb+Tucker+ materials, and by two \textsc{R} packages: \verb+WRI+~\citep{liu_wri_2022} and \verb+frechet+~\citep{chen_frechet_2023}. These packages face certain practical limitations. For instance, \verb+WRI+ requires continuous distributions, and does not allow user-specified constraints for the distribution support. \verb+frechet+ offers more flexibility in user specifications, but its solver for Fréchet regression is slow and may not accurately satisfy constraints.

The \verb+fastfrechet+ package addresses these limitations by providing a fast, scalable, and user-friendly implementation of both Fréchet regression and variable selection for 2-Wasserstein space, based on the work of
\citet{coulter_fast_2024}. The package incorporates resampling tools to enhance automatic variable selection, including cross-validation described in \citet{tucker_variable_2023} and stability selection described in \citet{coulter_fast_2024}. Additionally, \verb+fastfrechet+ features a customized dual active-set solver for the Fréchet regression problem, inspired by \citet{arnstrom_dual_2022}, which ensures both computational efficiency and accuracy while accommodating user-specified support constraints. The Fréchet regression solver also accommodates the auxiliary weighting scheme used in the variable selection procedure, the first package to
do so.

\section{Performance Comparisons to Existing Implementations}

We illustrate the performance of \verb+fastfrechet+ against existing implementations with simulated covariate-dependent distributional responses. The included function \verb+generate_zinbinom_qf+ simulates $n$ zero-inflated negative binomial (\textbf{zinbinom}) distributions (we choose $n = 100$), represented as quantile functions evaluated on a shared $m$-grid in $(0, 1)$ (we choose $m = 100$), and dependent on the first 4 of $p \geq 4$ covariates (we choose $p = 10$). We utilize the \textsc{R} package \verb+microbenchmark+~\citep{mersmann_microbenchmark_2024} to calculate run times, and report median times for each method (Fréchet regression, variable selection) from 15 iterations; all computations were performed on an Apple M1 Max chip. To replicate the specific simulation and comparison settings used in this manuscript, see the accompanying \verb+performanceExample-fastfrechet+ vignette.

\subsection{The Fréchet Regression Problem}

\verb+fastfrechet+ provides a solver for the Fréchet regression problem for
2-Wasserstein space, with optional \verb+lower+ and \verb+upper+ support constraints on
the underlying distributions. Since \textbf{zinbinom} distributions are non-negative,
we fix \verb+lower = 0+ and \verb+upper = Inf+ (or some suitably large number, as
applicable). The regression outputs are fitted quantile functions, which should
be monotone non-decreasing and obey support constraints. The \verb+fastfrechet+
implementation is a customization of the dual active-set method of~\citet{arnstrom_dual_2022}. (See the accompanying \verb+monotoneQP-fastfrechet+ vignette for
full algorithm description.)

\begin{figure}[!h]
    \centering
    \includegraphics[width=\linewidth]{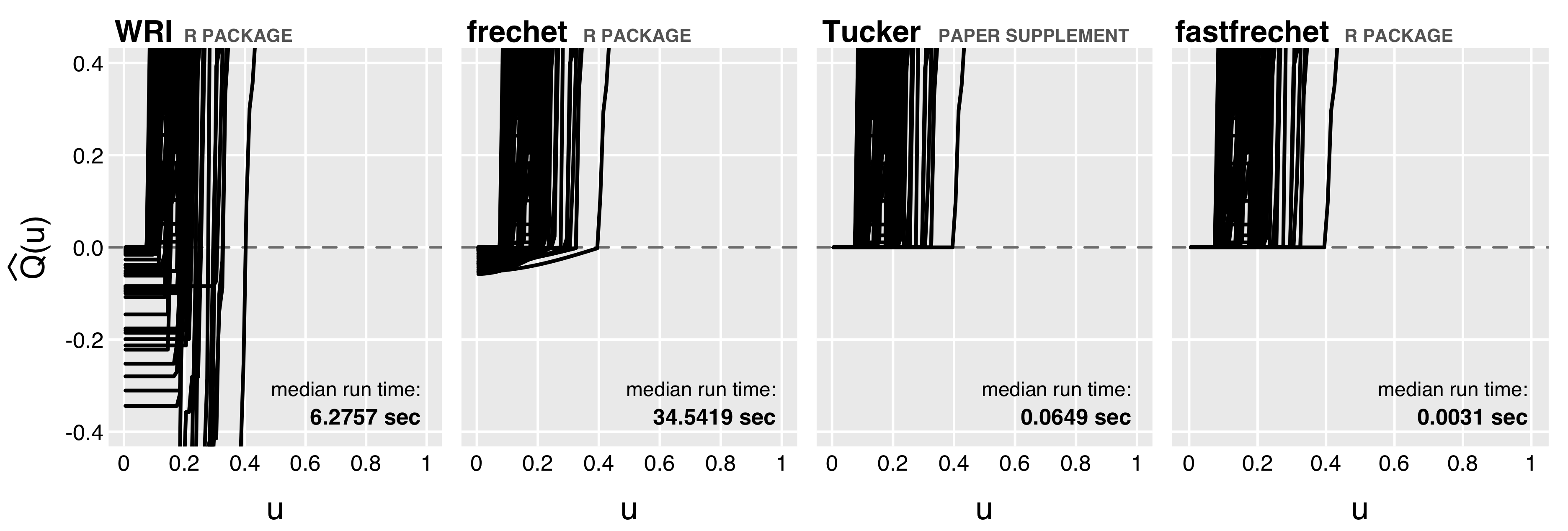}
    \cprotect\caption{Fitted Fréchet regression quantile functions (zoomed in around zero) and median run times for \verb+fastfrechet+ and other implementations. Fitted quantile functions below zero violate known lower support constraints.}\label{fig:frechetreg_comparison}
    \vspace{2em}
    \includegraphics[width=\linewidth]{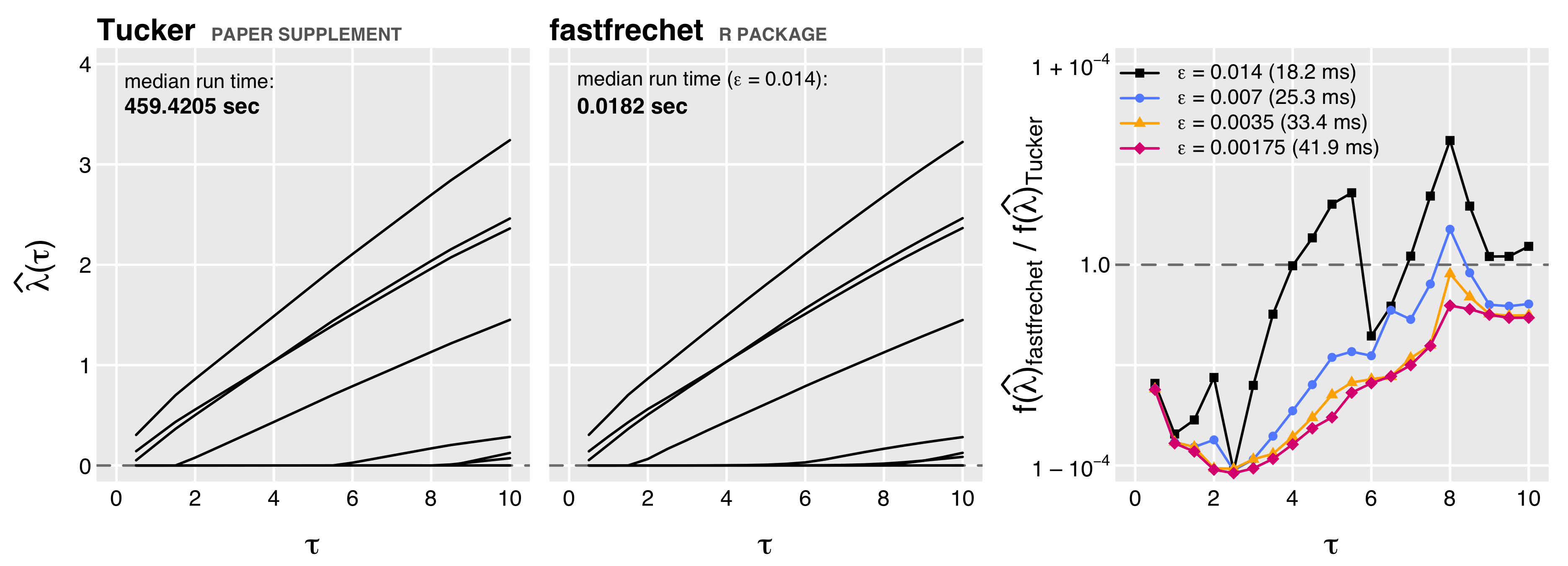}
    \cprotect\caption{(\textbf{left}, \textbf{center}) Variable selection solution paths $\widehat{\pmb{\lambda}}(\tau)$ across $\tau = \{0.5, 1, \dots, 10\}$ and median run times for \verb+Tucker+ materials and \verb+fastfrechet+. (\textbf{right}) Relative optimization accuracy of \verb+fastfrechet+ and \verb+Tucker+ materials variable selection, and median \verb+fastfrechet+ run times, using different error tolerance values. Points below 1.0 indicate \verb+fastfrechet+ solutions minimize the objective function better.}\label{fig:friso_comparison}
\end{figure}

Figure~\ref{fig:frechetreg_comparison} illustrates the speed and accuracy of
Fréchet regression implemented in \verb+fastfrechet+ against the \verb+WRI+, \verb+frechet+,
and \verb+Tucker+ materials implementations. \verb+WRI+ does not accept known support
bounds as input, and fitted responses correspondingly violate the zero lower
bound; \verb+frechet+ solutions only approximately satisfy the lower bound. The
\verb+Tucker+ materials implementation finds numerically accurate solutions, but
\verb+fastfrechet+ accomplishes this in a fraction the time.

\subsection{The Variable Selection Problem}

The R package \verb+fastfrechet+ implements variable selection for Fréchet
regression, specifically in 2-Wasserstein space. Variable selection comprises
finding optimal weight vector $\widehat{\pmb{\lambda}} \in \mathbb{R}^p$ that
satisfies a $\tau$-simplex constraint, given hyperparameter $\tau > 0$. In
2-Wasserstein space, $\widehat{\pmb{\lambda}}$ essentially minimizes an $L^2$
norm between weighted Fréchet regression outputs
$\widehat{\pmb{Q}}(\widehat{\pmb{\lambda}})$ and the raw data $\pmb{Y}$. (See
the accompanying \verb+intro-fastfrechet+ vignette for a detailed exposition.)
\verb+fastfrechet+ implements the second-order geodesic descent algorithm developed
by~\citet{coulter_fast_2024}, with two modifications. First, the implementation uses
the custom dual active-set method discussed in the previous subsection. The
active set defining the weighted Fréchet regression solution
$\widehat{\pmb{Q}}(\pmb{\lambda}^t)$ for iterate $\pmb{\lambda}^t$ serves as a
warm start for iterate $\pmb{\lambda}^{t + 1}$, reducing computation time.
Second, the implementation allows the user to specify an impulse parameter,
which implements momentum-based geodesic descent.

Figure~\ref{fig:friso_comparison} illustrates the speed and accuracy of variable
selection implemented in \verb+fastfrechet+ against the \verb+Tucker+ materials
implementation, across sequence of hyperparameter values
$\tau \in \{0.5, 1.0, \cdots, 10.0\}$. We hand-select \verb+fastfrechet+ error
tolerance parameter $\varepsilon = 0.014$, which gives solutions
$\widehat{\pmb{\lambda}}(\tau)$ minimizing the objective function
approximately as well as solutions from the other method ``as-is". \verb+fastfrechet+
is upward of 20,000$\times$ faster to obtain these comparable solutions.
Decreasing the \verb+fastfrechet+ error tolerance parameter increases optimization
accuracy with modest increases in computation time.

\section{Figures}

\clearpage
\bibliographystyle{chicago}
\bibliography{REFERENCES}

\end{document}